\documentclass[12pt]{article}
\newwrite\ffile\global\newcount\figno \global\figno=1

\def\writedef#1{}

\input epsf
\def\figin{\epsfcheck\figin}\def\figins{\epsfcheck\figins}
\def\epsfcheck{\ifx\epsfbox\UnDeFiNeD
\message{(NO epsf.tex, FIGURES WILL BE IGNORED)}
\gdef\figin##1{\vskip2in}\gdef\figins##1{\hskip.5in}
\else\message{(FIGURES WILL BE INCLUDED)}%
\gdef\figin##1{##1}\gdef\figins##1{##1}\fi}

\def\figinsert{}
\def\ifig#1#2#3{\xdef#1{fig.~\the\figno}
\writedef{#1\leftbracket fig.\noexpand~\the\figno}%
\figinsert\figin{\centerline{#3}}\medskip\centerline{\vbox{\baselineskip12pt
\advance\hsize by -1truein\center\footnotesize{  Fig.~\the\figno.} #2}}
\bigskip\endinsert\global\advance\figno by1}
\def\endinsert{}

\begin{document}
\baselineskip 18pt
\newcommand{\Tr}{\mbox{Tr\,}}
\newcommand{\beq}{\begin{equation}}
\newcommand{\eeq}{\end{equation}}
\newcommand{\bea}{\begin{eqnarray}}
\newcommand{\eea}[1]{\label{#1}\end{eqnarray}}
\renewcommand{\Re}{\mbox{Re}\,}
\renewcommand{\Im}{\mbox{Im}\,}
\begin{titlepage}

\begin{picture}(0,0)(0,0)
\put(350,0){SHEP-03-02}
\end{picture}

\begin{center}
\hfill
\vskip .4in
{\large\bf
The Yang Mills$^*$ Gravity Dual
}
\end{center}
\vskip .4in
\begin{center}
{\large David E. Crooks and Nick Evans}
\footnotetext{e-mail: dc@hep.phys.soton.ac.uk,
evans@phys.soton.ac.uk }
\vskip .1in
{\em Department of Physics, Southampton University, Southampton,
S017 1BJ, UK}

\end{center}
\vskip .4in
\begin{center} {\bf ABSTRACT} \end{center}
\begin{quotation}
\noindent We describe a ten dimensional supergravity geometry
which is dual to a gauge theory that is non-supersymmetric Yang Mills
in the infra-red but reverts to $N$=4 super Yang Mills in the
ultra-violet. A brane probe of the geometry shows that the scalar
potential of the gauge theory is stable. We discuss the infra-red behaviour
of the solution. The geometry describes a Schroedinger equation
potential that determines the glueball spectrum of the theory; there
is a mass gap and a discrete spectrum. The glueball mass predictions match
previous AdS/CFT Correspondence computations in the  non-supersymmetric
Yang Mills theory, and lattice data, at the 10$\%$ level. {\it Based on
a talk presented at SCGT02 in Nagoya, Japan.}
\end{quotation}
\vfill
\end{titlepage}
\eject
\noindent

\section{Introduction}

Dualities between gauge theories and string theories follow
naturally from the discovery of branes. The Born Infeld action for
the brane (like the Nambu-Goto string action) has a dual
interpretation as either describing a brane embedded in a
space-time or as a field theory living on the brane's surface. The
position of the brane in the bulk can equally be thought of as a
scalar vacuum expectation value in the field theory. The first
example of such a duality was the $AdS$/CFT Correspondence
\cite{mald} which is a duality between the conformal $N$=4 super
Yang Mills theory and IIB strings (supergravity) on five
dimensional Anti-de-Sitter space cross a five sphere. The field
theory's global symmetries (an SO(2,4) superconformal symmetry and
an SU(4)$_R$ symmetry) match to space-time symmetries of the $AdS$
space and the five sphere respectively. The supergravity fields
enter the field theory in symmetry invariant ways and so appear as
sources (eg masses) for field theory operators. The radial
direction in $AdS$ has the conformal symmetry properties of an
energy scale and has been interpreted as renormalization group
scale. Thus the radial behaviour of the supergravity fields
describes the RG flow of the field theory sources. Expectation
values of operators in the field theory are obtained from
derivatives with respect to these sources on the supergravity
partition function. The need to take derivatives suggests the
duality should hold in the presence of non-zero values for these
sources. We should therefore be able to study all possible
deformations of the $N$=4 super Yang Mills theory.

Techniques for introducing these deformations
\cite{gppz1,pilch,gppz3,freed2} and learning how to interpret them
\cite{gub,bpp,more} have  been developed over recent years. The
cleanest example \cite{freed2} involves the introduction of a vev
for the six adjoint scalar fields ($tr \phi_i \phi_j$) by allowing
a supergravity scalar field in the 20 representation of SU(4)$_R$
to be non-zero. Solutions of the five dimensional truncated
supergravity theory can be found but to interpret these geometries
they have been lifted to ten dimensions. In ten dimensions the
solutions can, for example, be brane probed \cite{more} and placed
in appropriate coordinates where they become multi-centre D3 brane
solutions. The original geometry was found from that around a
stack of D3 branes whose surface theory is the $N$=4 gauge theory.
Moving the branes apart, as in the multi-centre solutions, places
the theory on its moduli space and provides a natural gravity dual
in the presence of scalar vevs. The deformation program reproduced
these geometries and therefore seems to work well!

Here we will describe an on going attempt \cite{yms} to describe a
non-supersymmetric gauge theory using this technology. The four
adjoint fermions of the supersymmetric theory will be made massive
via a non-zero five dimensional supergravity field. The solution
will be lifted to a complete ten dimensional solution. Brane
probing then reveals the scalar potential and we will see that the
fermion mass radiatively generates a bounded mass for the six
scalar fields. The deep infra-red of this theory is therefore just
a gauge field. The ultra-violet theory is still the strongly
coupled and conformal $N$=4 theory. Since the UV is strongly
coupled there will never be a complete decoupling of the massive
matter fields from the dynamics. The goal is to find a theory with
the generic properties of QCD and only time will tell how good it
is as a numerical approximation. As a first step towards
uncovering the physics encoded by the geometry we study the
$O^{++}$ glueballs of the theory \cite{usgb}. The appropriate
Schroedinger equation potential \cite{gb,gub} is a bounded well
(providing further evidence of the stability of the solution) and
showing that there is both a mass gap and a discrete glueball
spectrum. We determine the spectrum and compare to the results
\cite{gb} from Witten's thermal AdS-Schwarzchild geometry
\cite{mald} and lattice simulations \cite{lat} of the
non-supersymmetric spectrum. Remarkably the results agree at the
10$\%$ level suggesting this approach may become a useful tool in
studying the non-supersymmetric theory.

\section{The Deformation in Five Dimensions}

We will introduce an equal  mass for the four adjoint fermions
of the $N$=4 theory via a five dimensional supergravity scalar in the
10 of SU(4)$_R$. The appropriate scalar, $\lambda$,
and its potential can be found in \cite{pilch}
($V = - {3 \over 2} \left[ 1 + \cosh^2 \lambda \right]$)
We look for solutions where $\lambda$ varies in the radial direction, $r$,
of $AdS$ and the metric is described by
\begin{equation}
ds_{(1,4)}^2 = e^{2 A(r)}dx^\mu dx_\mu + dr^2
\end{equation}
where $\mu=0..3$.

The equation of motion for the scalar fields are \cite{gppz1}

\begin{equation} \label{e1}
\lambda^{''} + 4 A^{'} \lambda^{'} = {\partial V \over \partial \lambda},
\hspace{1cm}
-3 A^{''} - 6 A^{'2} = \lambda^{'2} + 2 V
\end{equation}
Asymptotically, where the geometry returns to $AdS$, the solutions are
\begin{equation}
\lambda = M e^{-r} + K e^{-3r}
\end{equation}
Corresponding to a mass and a vev for our fermionic operator.

\begin{figure}[ht]
\centerline{\epsfxsize=3.5in\epsfbox{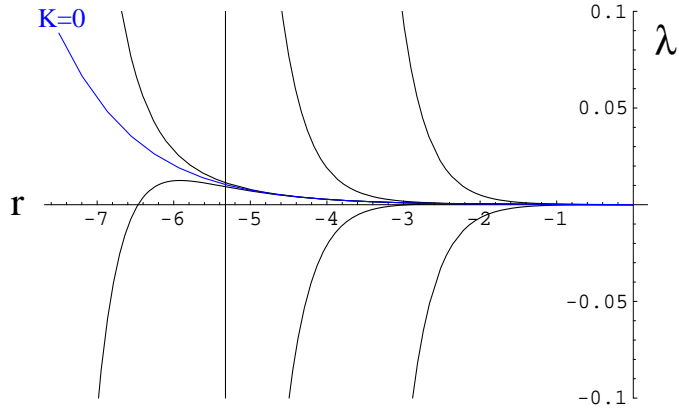}}
\caption{Numerical solutions of the 5d supergravity equations for the
scalar $\lambda$. The $K$=0 flow corresponds to the mass only boundary
conditions.}
\end{figure}

Numerical solution of these equations are displayed in figure 1
for different asymptotic boundary conditions. The mass only flow
is a unique flow - in the presence of any condensate the flows
clearly diverge. Finding the final fate of the mass only flow
numerically requires arbitrary fine tuning of the initial
conditions. However, from figure 1 it seems likely the flow
diverges in the very deep infra-red. The interpretation of such
singularities remains open. For example the backgrounds describing
$N$=4 SYM on moduli space  are singular but those singularities
are understood to correspond to the presence of D3 branes in the
solution. In the $N$=2$^*$ theory \cite{bpp} the singularities
correspond to the divergence of the running gauge coupling. This
latter case is the most likely explanation of the divergence here.
We will see that a well defined glueball spectrum emerges from
this geometry in spite of the divergence suggesting it is not a
disaster!

Interpreting the five dimensional geometries has proven hard so we will
move to the lift of the solution to ten dimensions.

\section{The Ten Dimensional Lift}

Lifting the five dimensional solutions to ten seems like a tough task
but Pilch and Warner have made an ansatz \cite{pilch}
for the form of the metric and
dilaton. The remaining ten dimensional forms can then be found from
the equations of motion (after much work!). The lift is described in
detail in \cite{yms} but here we will just present the results.

Asymptotically the scalar in the 10 lifts to a 3-form potential
\begin{equation}
A_{(2)} = 2 \lambda (i \cos^3 \alpha \cos \theta_+
d \theta_+ \wedge d \phi_+
- \sin^3 \alpha  \cos \theta_-  d \theta_- \wedge d \phi_-)
\end{equation}
We have written the five-sphere as two 2-spheres ($d \Omega_\pm^2 =
d \theta_\pm^2 + \sin^2 \theta_\pm d \phi_\pm^2$) and an angle $\alpha$
between them.

The full solution has all the ten dimensional fields switch on. The metric
is given by

\begin{equation}
ds_{10}^2 = (\xi_+ \xi_-)^{1 \over 2} ds_{1,4}^2+
(\xi_+ \xi_-)^{-{3 \over 2}}  ds_{5}^2
\end{equation}

\begin{equation}
ds_5^2 =\xi_-  \cos^2 \alpha  ~ d\Omega_{+}^2
+ \xi_+ \sin^2 \alpha  ~ d\Omega_{-}^2
+\xi_+ \xi_- d\alpha^2
\end{equation}
where the $\xi_\pm$ are given by

\begin{equation}
\xi_\pm = c^2 \pm s^2 \cos 2\alpha,
\hspace{0.5cm} c= \cosh \lambda, \hspace{0.5cm}
s = \sinh \lambda
\end{equation}

The dilaton is given, in unitary gauge, by the functions

\begin{equation}
f = { 1 \over \xi^{1/2}} \sqrt{\cosh^2 \lambda + (\xi_+ \xi_-)^{1/2}
\over 2}, \hspace{1cm}
B = {  \sinh^2 \lambda \cos 2 \alpha \over  \cosh^2 \lambda
+ (\xi_+ \xi_-)^{1/2}}
\end{equation}
In the more usual language the axion-dilaton field is given by

\begin{equation}
C + i e^{-\Phi} = i { ( 1 - B ) \over (1 + B)}
\end{equation}
Thus for this solution the $\theta$ angle is switched off.

The two-form potential is given by

\begin{equation}
A_{(2)}=i A_{+} \cos^3 \alpha \, \cos \theta_{+}
d \theta_{+} \wedge d \phi_{+}
-  A_{-}
\sin^3 \alpha \, \cos \theta_{-} d \theta_{-} \wedge d \phi_{-}
\end{equation}
with

\begin{equation}
A_{\pm}=\sinh 2\,\lambda / \xi_\pm
\end{equation}
Finally the four-form potential lifts to
\begin{equation}
F_{(4)}=  F +\star F, \hspace{0.5cm} F
= dx^{0}\wedge dx^{1}\wedge dx^{2}\wedge dx^{3}\wedge d\omega
\end{equation}
where

\begin{equation}
\omega(r)=e^{4A(r)} A'(r)
\end{equation}

\section{Brane Probing}

As a first exploration of this geometry we can place a probe D3 brane
in the geometry. At leading order in $1/N$ we can neglect the back reaction
of the probe on the geometry. Substituting the geometry into the Born
Infeld action for the probe
\begin{equation} \label{BI}
S_{probe}=-\tau_3\int_{\mathcal{M}_4}d^4x
\det[G^{(E)}_{ab} + 2 \pi \alpha' e^{- \Phi/2} F_{ab}]^{1/2}
+ \mu_3\int_{\mathcal{M}_4} C_4,
\end{equation}
will reveal the field theory on the brane's surface. We find a potential

\begin{equation}
V_{probe} = e^{4A} \left[\xi_+ \xi_-
- A^{'} \right]
\end{equation}
It is illuminating to evaluate this potential at leading order in the
ultra-violet with $\lambda = M  e^{- r}$, $A = r$, which gives

\begin{equation}
V = M^2 e^{2 r} + ...
\end{equation}
The field $e^r$ has conformal dimension 1 and should be identified
with the scalar fields of the field theory. This term corresponds
to an equal, bounded mass for the six scalars. This confirms field
theory expectations that when supersymmetry is broken via the
fermion mass the scalars will radiatively acquire a mass. It is
also encouraging that the theory has a bounded potential.

\section{The Glueball Spectrum}

We can make an initial investigation of the infra-red properties
of the gauge theory described by our geometry as follows. The
$O^{++}$ glueballs of the theory have been identified \cite{gb}
with excitations of the dilaton field of the form
\begin{equation}
\delta \Phi = \psi(r) e^{-i k x}, \hspace{1cm} k^2 = -M^2
\end{equation}

\noindent This deformation must be a solution of the 5d dilaton field
equation $\partial_\mu( \sqrt{-g} g^{\mu \nu} \partial_\nu) \delta \Phi
= 0$. If we make the change of coordinates \cite{gub}
($r \rightarrow z$) such that

\begin{equation} {d z \over dr} = e^{2 A}, \hspace{1cm}
\psi \rightarrow e^{-3 A/2} \psi\end{equation}
Then the dilaton field equation takes a Schroedinger form

\begin{equation} ( - \partial_z^2 + V(z)) \psi(z) = M^2 \psi(z)  \end{equation}
where
\begin{equation}
V= {3 \over 2} A^{''} + {9 \over 4} (A^{'})^2
\end{equation}

Solving the equations of motion in these coordinates and tuning onto the
mass only solution produces the well potential shown in figure 2 \cite{usgb}.
Note
that if any condensate is present the

\begin{figure}[ht]
\centerline{\epsfxsize=3.5in\epsfbox{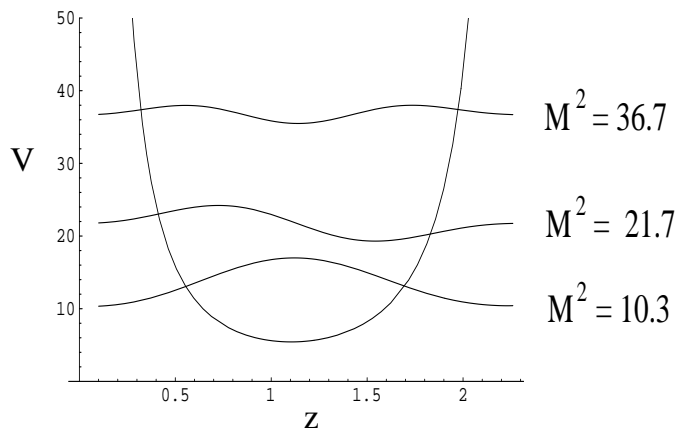}}
\caption{The Schroedinger potential for the $O^{++}$ glueballs and the lowest
lying solutions.}
\end{figure}

\noindent well becomes unstable at large $z$.
In the massive case the potential well shows us that there is a mass gap and
a discrete glueball spectrum. The gauge theory dual is confining in the
infra-red.

The glueball spectrum can be obtain using the numerical shooting technique
and the three lowest energy solutions are shown in figure 2. We therefore have
predictions for the lightest $O^{++}$ glueball states, shown in table 1.
The lightest state's mass is not a prediction but can be used to fix the value
of $\Lambda_{QCD}$ - we normalize it to the lattice results discussed below.
It is interesting
to compare to other computations of these masses. Witten \cite{mald}
found a high
temperature deformation of the gravity dual of the field theory
on the surface of an M5 brane
which is expected at low energies to describe 4 dimensional non-supersymmetric
Yang Mills theory (but in the UV
has many extra adjoint matter fields and lives in 6 dimensions).
Similar techniques were used
to determine the predicted glueball masses \cite{gb} and are shown in Table 1.
They match remarkably well with our results suggesting that the high energy
completion of the theory is relatively unimportant. We also display the limited
lattice results in non-supersymmetric Yang Mills in the table and again the
agreement is at the 10$\%$ level although we only have the one excited state
result for comparison. \bigskip

\begin{center}
\begin{tabular}{|c|c|c|c|}
\hline
& YM$^*$ & AdS-Schwarz & Lattice\\
\hline
O$^{++}$ & 1.6 (input) & 1.6 (input) & $1.6 \pm 0.15$\\
O$^{++*}$ & 2.4 & 2.6 & $2.48 \pm 0.23$\\
O$^{++**}$ & 3.1 & 3.5 & ? \\
\hline
\end{tabular}\label{tab1} \bigskip

Table 1: Glueball mass predictions from the $AdS$/CFT
Correspondence and lattice calculations.\vspace*{1pt}
\vspace*{-13pt}
\end{center} \bigskip

Encouragingly  the Yang Mills$^*$ gravity dual appears to encode
much of the physics we would expect of non-supersymmetric Yang
Mills theory. The obvious next challenge is to include quark
fields which we are currently working on.

\section{Acknowledgements}

DEC is grateful to PPARC for the support of a studentship. NE is grateful
to PPARC for the support of an Advanced Fellowship.

\end{document}